\documentclass[12pt]{article}

\usepackage{graphicx,capt-of}
\usepackage{color}
\usepackage{amssymb,amsmath,amsfonts,bm}
\usepackage{placeins}
\usepackage{subfigure}
\usepackage{caption}
\usepackage{tabularx,ragged2e,booktabs,caption}
\usepackage{grffile}
\makeatletter
\newcommand*{\rom}[1]{\expandafter\@slowromancap\romannumeral #1@}
\makeatother

\def\beq{\begin{equation}}
\def\eeq{\end{equation}}
\def\bea{\begin{eqnarray}}
\def\eea{\end{eqnarray}}

\def\la{\langle}
\def\ra{\rangle}

\begin{document}

\begin{center}
{\Large{\bf Phase Ordering Kinetics of the Asymmetric Coulomb Glass Model}} \\
\ \\
\ \\
by \\
Preeti Bhandari$^{1}$, Vikas Malik$^2$ and Sanjay Puri$^3$ \\
$^1$Department of Physics, Ben Gurion University of the Negev, Beer Sheva 84105, Israel. \\
$^2$Department of Physics and Material Science, Jaypee Institute of Information Technology, Uttar Pradesh 201309, India. \\
$^3$School of Physical Sciences, Jawaharlal Nehru University, New Delhi -- 110067, India.
\end{center}

\begin{abstract}
We present results for phase ordering kinetics in the {\it Coulomb glass} (CG) model, which describes electrons on a lattice with unscreened Coulombic repulsion. The filling factor is denoted by $K \in [0,1]$. For a square lattice with $K=0.5$ (symmetric CG), the ground state is a checkerboard with alternating electrons and holes. In this paper, we focus on the asymmetric CG where $K \lesssim 0.5$, i.e., the ground state is checkerboard-like with excess holes distributed uniformly. There is no explicit quenched disorder in our system, though the Coulombic interaction gives rise to frustration. We find that the evolution morphology is in the same dynamical universality class as the ordering ferromagnet. Further, the domain growth law is slightly slower than the {\it Lifshitz-Cahn-Allen} law, $L(t) \sim t^{1/2}$, i.e., the growth exponent is underestimated. We speculate that this could be a signature of logarithmic growth in the asymptotic regime.
\end{abstract}

\newpage

\section{Introduction}
\label{s1}

Many systems found in nature are far from equilibrium. We still lack a complete understanding of non-equilibrium pattern formation in these systems. An important problem in this context is the temporal evolution of a system which is rendered thermodynamically unstable by a rapid quench below the critical temperature $T_c$. The system does not instantaneously transform to its new phase -- rather, but small domains of coexisting phases grow with time to reach the equilibrium state. This evolution is usually referred to as {\it phase ordering dynamics} or {\it domain growth} or {\it coarsening} \cite{pw09,dp04}. We have a good understanding of domain growth in pure and isotropic systems. The nature of evolution towards equilibrium depends on whether or not the order parameter is conserved, and the nature of defects (interfaces, vortices, monopoles, etc.) which drive the coarsening process. Typically, non-conserved systems with short-ranged interactions are characterized by the {\it Lifshitz-Cahn-Allen} (LCA) growth law for the domain size $L$ at time $t$: $L(t) \sim t^{1/2}$ \cite{pw09,dp04}.

Recent interest in domain growth problems has focused on experimentally realistic systems. Most experimental systems have disorder and frustration. Therefore, it is relevant to study the ordering kinetics of such systems. In a system with quenched disorder, the dynamics of the system slows down due to the trapping of domain walls by disordered sites \cite{hh85,lmv88,sp04}. Another important system property that affects the growth is the range of the particle-particle interaction. There have been several analytical \cite{br94} and numerical \cite{acl21,cmj21} studies of domain growth in systems with long-range interactions. These have demonstrated that the growth law is modified from the LCA law if the interactions are sufficiently long-ranged.

In this paper, we study coarsening in an important model which combines disorder/frustration with long-range interactions. We focus on the well-known Coulomb glass (CG) model \cite{es75,se84} at half-filling ($K = 0.5$), and also at fillings slightly less than half ($K=0.475, 0.45$). In the CG model, the electronic states are localized due to disorder. Due to the localization of states, the electron-electron interaction is unscreened. The Hamiltonian for the CG model is
\begin{equation}
\label{HamiltonianP}
\mathcal{H}_{\rm CG} \{n_i\} = \sum_{i=1}^N (n_i - K) \phi_i + \frac{1}{2} \sum_{i \neq j} \frac{e^{2}}{\kappa |\vec{r_i} - \vec{r_j}|} \left( n_i-K \right)\left(n_j-K \right) ,
\end{equation}
where $N$ is the number of lattice sites, and $n_{i}$ $\in \{0,1\}$ is the occupation number of the electron on site $i$ with position vector $\vec{r}_i$. In Eq.~(\ref{HamiltonianP}), $\phi_{i}$ is the random energy at $i$, and $K$ is the filling factor. The charge on each electron is equal to $-e$ and a compensating charge of $+K e$ is put on each site to maintain charge neutrality. The second term in Eq.~(\ref{HamiltonianP}) denotes an unscreened Coulombic interaction between the charges at sites $i$ and $j$ in a medium of dielectric constant $\kappa$.

At high disorder and low temperature, experiments have shown that this system exhibits glassy behavior such as aging \cite{vop00,tg04,zo06,stp14}, slow relaxation of excess conductance \cite{bop93,mcg98,tg03}, and memory effects \cite{vop02,lm05}. Recent simulations of the $d=3$ CG via {\it population annealing} have shown a transition from a disordered to a glassy phase as the temperature is lowered at high values of disorder \cite{bas19}. No glassy phase was found in earlier {\it exchange Monte Carlo} (MC) simulations done by Goethe and Palassini \cite{gp09} and Surer et al. \cite{skz09}. The mean-field study of the model also indicates the presence of a glassy phase \cite{pd99,pd05,mp07}.

The CG model is equivalent to an Ising anti-ferromagnet (AF) with spins interacting via the Coulombic interaction. In $d=3$, equilibrium studies at half-filling (symmetric CG) have shown that the CG is in the same universality class as the {\it random field Ising model} (RFIM) with ferromagnetic nearest-neighbor (nn) interactions \cite{gp09}. The system undergoes a second-order phase transition from a paramagnetic phase to an AF phase at zero and finite disorders. At zero disorder, the critical exponents of the CG match those of the $d=3$ nn Ising model \cite{mr09,bm20}. For disorder amplitudes less than the critical disorder, the critical exponents are those of the $d=3$ nn RFIM.

In $d=2$, at zero disorder, it has been found that the critical exponents of the CG match those of the $d=2$ nn Ising model \cite{mr09,bm19}. However, unlike the $d=2$ nn RFIM, the CG shows a transition from a disordered to a charge-ordered phase at $T=0$ as the disorder becomes less than a critical value \cite{bma17}.

To the best of our knowledge, there have been far fewer studies of dynamical properties of the CG. In recent work \cite{bmp19}, we investigated the ordering kinetics of the $d=2$ symmetric ($K=0.5$) CG at zero and small disorder. In conformity with experiments, the composition of the CG is conserved during the evolution. However, the relevant order parameter ({\it staggered magnetization}) is a nonconserved variable \cite{ddp17}. For zero disorder, the CG  system obeyed the LCA growth law \cite{bmp19}. Thus, in spite of frustration due to the long-range Coulomb interaction, the CG kinetics is the same as that of the nn AF. At non-zero disorder, the system showed a transition from power law to logarithmic growth similar to that observed in the nn RFIM \cite{clm12}.

In this paper, we focus on the effect of composition asymmetry on the ordering kinetics of the CG, i.e., we consider filling fractions away from $K=0.5$. We point out that the Hamiltonian in Eq.~(\ref{HamiltonianP}) is invariant under the transformation $K \rightarrow 1-K$ and $n_i \rightarrow 1-n_i$. Thus, the case with surplus electrons ($K > 0.5$) is equivalent to the case with surplus holes with composition $1-K$. Therefore, without loss of generality, we consider the case with $K \lesssim 0.5$. For simplicity, we consider the case with no external disorder. The internal frustration of the system already yields rich phenomenology, which we will discuss in this paper.

This paper is organized as follows. In Sec.~\ref{s2}, we give details of our numerical simulations. In Sec.~\ref{s3}, we present detailed numerical results. In Sec.~\ref{s4}, we conclude this paper with a summary and discussion.

\section{Simulation Details}
\label{s2}

Recently, Rademaker et al. \cite{rpz13} used mean-field theory to obtain a finite temperature phase diagram for the $d=2$ CG model over a range of average particle density. Using their phase diagram as a reference point, we have studied ordering kinetics at $K=0.45,0.475,0.50$. The first two compositions differ only slightly from half-filling, and we assume that the critical temperature ($T_c$) is approximately the same for all three $K$-values. From the phase diagram proposed by Rademaker et al., one can see that the ground state has checkerboard order close to half-filling. This implies that one can still use staggered magnetization as an order parameter for $K=0.45 \rightarrow 0.5$. For $K \simeq 0.4$, the phase diagram shows that the ground state has a stripe order. The ground states become more exotic as $K$ is reduced further. For this reason, we have chosen $K=0.45$ as the lower limit of asymmetry for our simulation.

Before proceeding further, it is useful to formulate the CG Hamiltonian in spin language. We introduce the usual spin notation: $S_i = 2 n_i-1$. Then, Eq.~(\ref{HamiltonianP}) with $\phi_i = 0$ can be written as
\begin{equation}
\label{Hamiltonian}
\mathcal{H}_{\rm CG} \{S_i\} = \frac{1}{8} \sum_{i \neq j} \frac{e^{2}}{\kappa |\vec{r_i} - \vec{r_j}|} (S_i+1-2K)(S_j+1-2K), \quad S_i = \pm 1 .
\end{equation}
For half-filling ($K=0.5$), the constant factor $1-2K=0$. We would like to compare domain growth in the $d=2$ CG with a system in which the Coulombic interaction is screened, i.e., charges (spins) interact via a short-range potential. The latter is modeled by a nn AF with charge neutrality. The Hamiltonian can be written as
\begin{equation}
\label{Hamiltonianising}
\mathcal{H}_{\rm AF} \{S_i\} = J \sum_{\la ij \ra} (S_i+1-2K)(S_j+1-2K) , \quad S_i = \pm 1 ,
\end{equation}
where $J>0$ is the strength of the nn interaction, and $\la ij \ra$ denotes nn pairs. We will subsequently use the term ``CG'' to refer to the model in Eq.~(\ref{Hamiltonian}), and the term ``AF'' to refer to the model in Eq.~(\ref{Hamiltonianising}).

The details of our simulation are as follows. We study coarsening via Monte Carlo (MC) simulations \cite{bh02,nb99} of the CG and AF with Kawasaki spin-exchange kinetics. (We will shortly discuss this kinetics in detail.) We use a square lattice of size $N = A^2$, and employ periodic boundary conditions in both directions. To cope with the long-range Coulomb interactions in Eq.~(\ref{Hamiltonian}), we have used the Ewald summation technique \cite{fs96}. The simulation of long-range interactions is computationally much more demanding than that of nn interactions. This restricts the size of the CG to $A=512$. The corresponding size of the AF is $A=2048$.

The initial configuration was prepared by randomly assigning $K N$ sites with value  $S_i = +1$, and $(1-K) N$ sites with $S_i = -1$. This mimics the disordered state at high $T$. The system was rapidly quenched at time $t=0$ from the disordered state ($T \gg T_{c}$) to a low-$T$ state ($T \sim T_{c}/2$). As the number of electrons in the system is conserved, the stochastic kinetics used to study the ordering process is Kawasaki spin-exchange kinetics \cite{pw09}. A pair of opposite spins $S_n$ and $S_m$ (which are nearest neighbors) are chosen at random for spin exchange. This exchange is done with probability $p$, given by the Metropolis algorithm \cite{bh02,nb99}:
\bea
\label{MC_prob}
p &=& 1, \quad \Delta_{nm} \leqslant 0 , \nonumber \\
&=& e^{-\beta \Delta_{nm}}, \quad \Delta_{nm} > 0 ,
\eea
where $\Delta_{nm}$ denotes the change in energy due to the spin exchange, and $\beta = 1/(k_B T)$ (Boltzmann constant $k_B = 1$). The energy change for the CG can be computed from Eq.~(\ref{Hamiltonian}) as follows:
\beq
\label{deltanm}
\Delta^{\rm CG}_{nm} = \frac{1}{4} (S_m-S_n) (\varepsilon_n - \varepsilon_m) - \frac{e^2}{\kappa r_{nm}} .
\eeq
In Eq.~(\ref{deltanm}), $\varepsilon_n$ denotes the single-particle Hartree energy:
\beq
\varepsilon_n = \sum_{j \neq n} \frac{e^2}{\kappa r_{nj}}~(S_j+1-2K) .
\eeq
The $\varepsilon_i$'s are computed at the beginning of the simulation and updated as the spin exchanges take place. The advantage of using the $\varepsilon_n$'s is that they are updated only when a spin exchange attempt is successful. This standard method for the long-range model significantly decreases the computational time needed for an update. A single Monte Carlo step (MCS) corresponds to $N$ attempted updates.

The corresponding expression for $\Delta_{nm}$ for the AF is considerably simpler, as expected. We use Eq.~(\ref{Hamiltonianising}) to obtain
\beq
\label{deltaf}
\Delta^{\rm AF}_{nm} = J (S_m-S_n) \left( \sum_{L_n \neq m} S_{L_n} - \sum_{L_m \neq n} S_{L_m} \right) ,
\eeq
where $L_n$ refers to the nn of site $n$.

All the statistical data presented in this paper has been averaged over 20 independent runs for the CG. The energies and temperatures of the CG system are calculated in units of $e^{2}/(\kappa a)$, where $a$ is the lattice constant. The critical temperature for the CG at half-filling \cite{mr09,bm19} is $T_c \simeq 0.1042~e^2/(\kappa a)$.

For the AF, the statistical results were obtained by averaging over 10 independent runs. In this case, the unit of energy is $J$, and $T_c \simeq 2.269~J$. 

\section{Detailed Numerical Results}
\label{s3}

\begin{figure}
\centering
\includegraphics[width=0.9\textwidth]{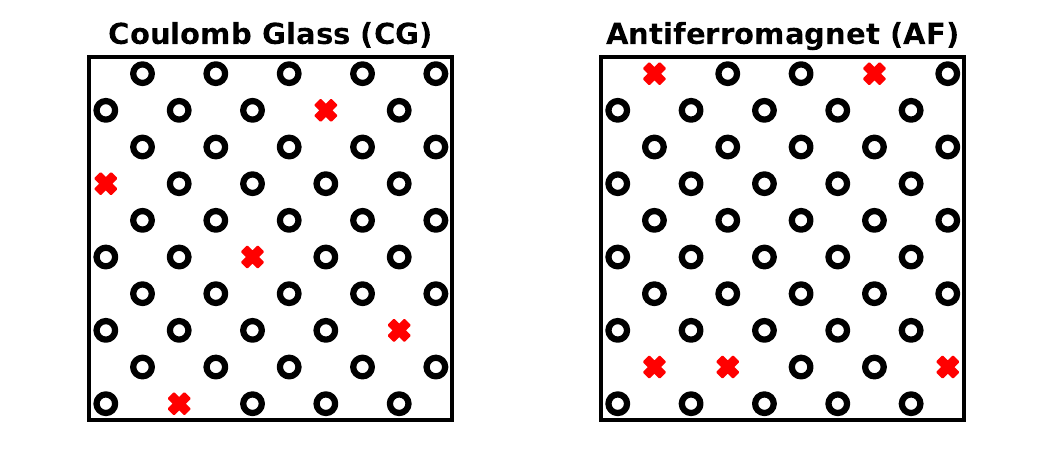}
\caption{Schematic of ground states of the (a) CG and (b) AF with slightly asymmetric composition: $K \lesssim 0.5$. The circles denote electrons, and the bars denote holes. The surplus holes are colored red.}
\label{f1}
\end{figure}

In earlier work, we have established that both the CG and AF in $d=2$ at $K = 0.50$ follow the LCA growth law \cite{bmp19}. We now investigate the role of excess electrons/holes or asymmetry in ordering kinetics. Without loss of generality, we consider the case with excess holes ($K < 0.5$). As discussed earlier, we only consider the slightly asymmetric case ($K > 0.45$), so that the system still has a checkerboard-like ground state. The excess holes are distributed uniformly throughout the system in both the CG and AF. These two situations are shown schematically in Fig.~\ref{f1}. In the CG, the excess holes form a periodic structure to minimize the Coulombic energy. No such periodicity is seen in the AF. The presence of excess holes at the domain interfaces slows down domain growth for both models. As mentioned earlier, the appropriate order parameter in this problem is not the composition (which is conserved) but rather the staggered magnetization (which is not conserved):
\beq
\sigma_i = (-1)^{i_x+i_y} S_i .
\eeq
The conserved composition plays the role of an auxiliary slow variable. A complete description of the kinetics is provided by {\it Model C} in the terminology of Hohenberg-Halperin \cite{hh77,ddp17}.

\begin{figure}
\centering
\includegraphics[width=0.95\textwidth]{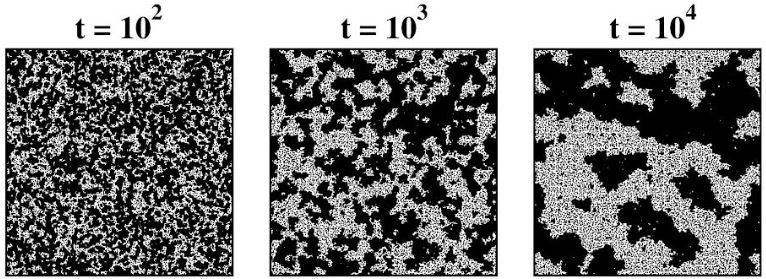}
\caption{Domain growth in the $d = 2$ CG at $K = 0.45$. We show snapshots at different MCS for a $512^{2}$ lattice. The order parameter is the staggered magnetization $\sigma$. The black and white regions correspond to $\sigma > 0$ and $\sigma < 0$, respectively.}
\label{f2}
\end{figure}

\begin{figure}
\centering
\includegraphics[width=0.95\textwidth]{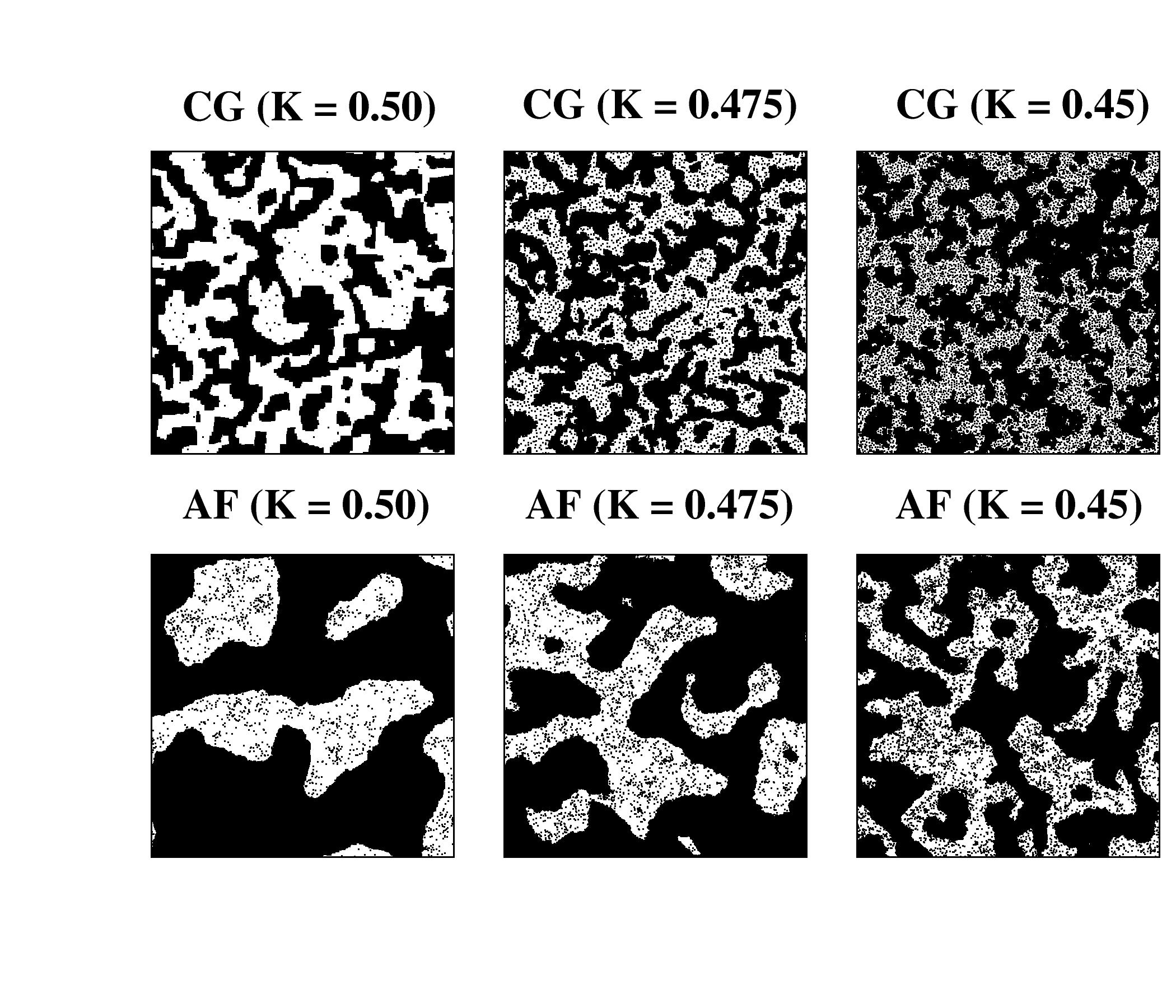}
\caption{The upper frames show snapshots at $t = 10^{3}$ MCS from domain growth in the $d = 2$ CG of size $512^2$. We show snapshots at different filling factors $K$. The order parameter is the staggered magnetization $\sigma$. The lower frames show corresponding snapshots for the AF. These frames show a $512^2$ corner of a $2048^2$ lattice.} 
\label{f3}
\end{figure}

In Fig.~\ref{f2}, we present evolution snapshots of $\sigma$-domains at $K = 0.45$ for the CG. The excess holes, which can correspond to either $\sigma_i = +1$ or $\sigma_i = -1$ (depending on the location) are seen to be uniformly dispersed as impurities in the domains of Fig.~\ref{f2}. In Fig.~\ref{f3}, we compare domain morphologies at different filling factors (at $t = 10^{3}$ MCS) for the CG and AF. For both models, one sees that domain growth slows down as the number of excess holes or domain impurities increases. (Notice that the snapshots for $K=0.50$ do not show a perfect checkerboard structure either -- this is a consequence of thermal fluctuations in the equilibrated bulk domains.)
 
The phase ordering kinetics in a system is usually characterized by the equal-time correlation function \cite{pw09}:
\beq
\label{Correlation}
C(\vec{r},t) = \frac{1}{N} \sum_{\vec{R}} \left[ \la \sigma(\vec{R},t) \sigma(\vec{R}+\vec{r},t) \ra - \la \sigma(\vec{R},t) \ra \la \sigma(\vec{R}+\vec{r},t) \ra \right] ,
\eeq
where we have assumed that the system is translationally invariant. This quantity measures the overlap of staggered magnetization between two spins separated by $\vec{r}$. Here, $\la \cdots \ra$ indicates an average over independent runs, i.e., with different initial conditions and noise realizations. If the system is isotropic and is characterized by a single length scale $L(t)$, $C(\vec{r},t)$ obeys the following {\it dynamical scaling} form \cite{pw09}:
\begin{equation}
\label{Corr_scale}
C(\vec{r},t) \equiv C(r,t) = f \bigg( \frac{r}{L(t)}\bigg) ,
\end{equation}
where $r$ is the magnitude of $\vec{r}$.

\begin{figure}
\centering
\includegraphics[width=0.8\textwidth]{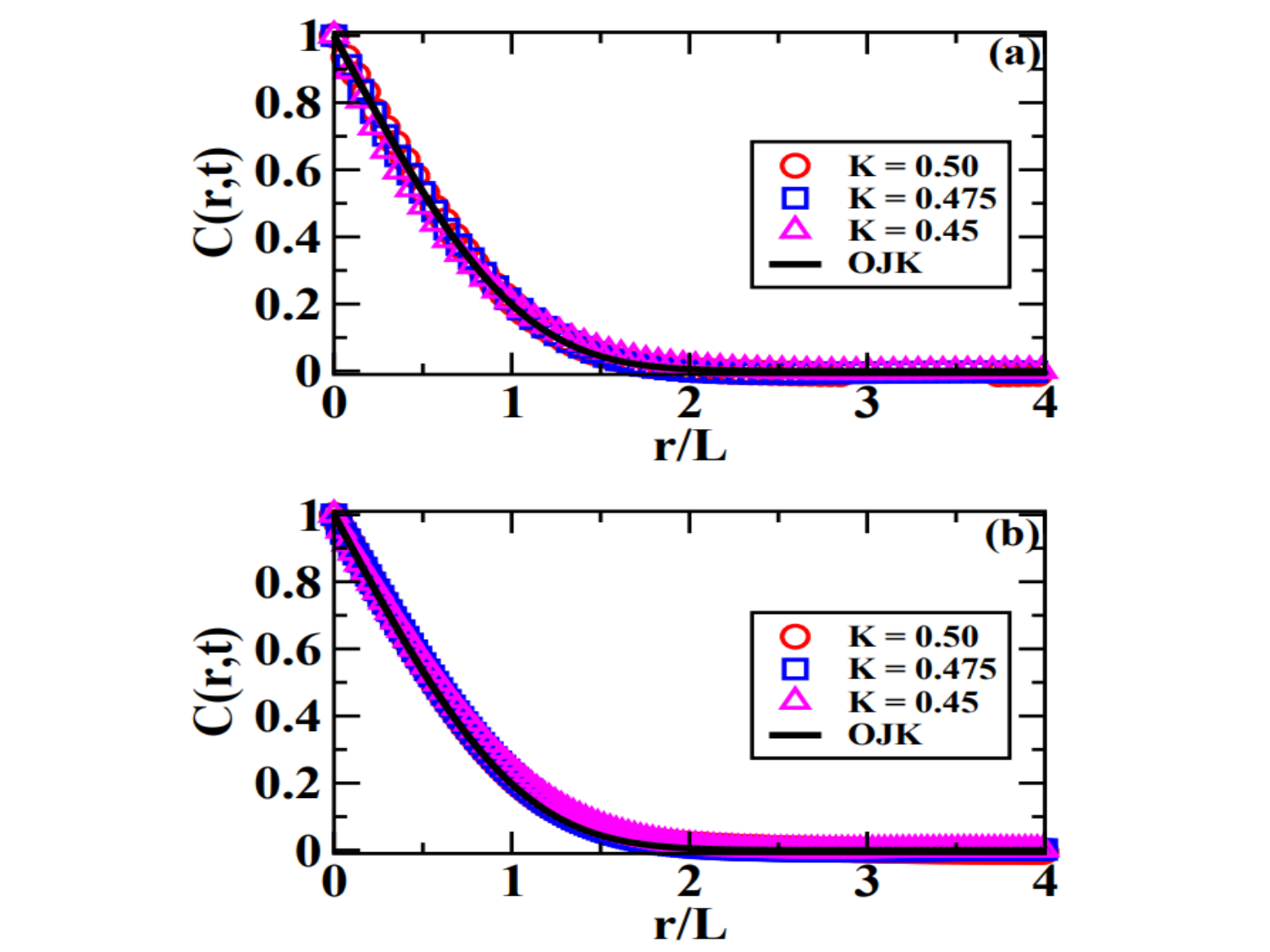}
\caption{(a) Scaled correlation functions, $C(r,t)$ vs. $r/L$, for the CG at $t=10^{3}$ MCS and different $K$. The length scale $L(t)$ is defined as the distance over which $C(r,t)$ decays to 0.2 of its maximum value. (b) Analogous plot for the AF. The solid line in both frames denotes the OJK function in Eq.~(\ref{ojk}).}
\label{f4}
\end{figure}

The characteristic domain size $L(t)$ is defined as the distance over which $C(r,t)$ decays to a fraction (say 0.2) of its maximum value. The scaling property indicates that the morphology is statistically self-similar in time -- only the scale of the morphology changes. We have confirmed (not shown here) that dynamical scaling holds for both the CG and AF. In Fig.~\ref{f4}(a), we plot the scaling functions for the CG with different values of $K$. The data sets are seen to be in excellent agreement. The solid line denotes the {\it Ohta-Jasnow-Kawasaki} (OJK) function \cite{pw09}:
\beq
\label{ojk}
f_{\rm OJK} (r/L) = \frac{2}{\pi} \sin^{-1} \left[ \exp (-r^2/L^2) \right] .
\eeq
The OJK function is an approximate result for the scaling function of the nn ferromagnet with nonconserved kinetics. It is obtained by modeling the pattern dynamics via the motion of domain boundaries or defects \cite{pw09}. We see that the OJK function describes the CG ordering kinetics very well, at least for the small values of asymmetry considered here. In Fig.~\ref{f4}(b), we show the scaling functions for the AF with small asymmetry. These are also in excellent agreement with the OJK function.

Next, let us investigate the time-dependence of the domain growth law. Typically, the evolution of domains in magnetic ordering is curvature-driven, and is governed by the equation \cite{lmv88,sp04}
\begin{equation}
\label{growthEq}
\frac{dL}{dt} = \frac{a(L,T)}{L} .
\end{equation}
Here, $a(L,T)$ is the diffusion constant which, in general, depends upon the domain size and temperature. Lai et al. (LMV) \cite{lmv88} classified systems with nonconserved kinetics via the functional dependence of $a(L,T)$. Consider the situation where coarsening domains of size $L$ encounter energy barriers $E_B(L)$. Then, the growth dynamics is governed by thermally-activated hopping over these barriers, and the diffusion constant has the form ($k_B=1$)
\begin{equation}
\label{diff_const}
a(L,T)= a_0~\exp \left(-\frac{E_{B}(L)}{T} \right) ,
\end{equation}
where $a_0$ is a constant. For the usual nn Ising model with {\it Glauber spin-flip} kinetics \cite{pw09}, there are no energy barriers ($E_B=0$) and $a(L,T)$ is a constant. Then, Eq.~(\ref{growthEq}) yields the LCA growth law:
\begin{equation}
\label{LCA}
L= (2 a_0 t)^{1/2} .
\end{equation}
These are designated as {\it Class 1} systems by LMV. For {\it Class 2} systems, the barrier energy is non-zero but independent of $L$: $E_B(L) \equiv E_B$. In that case, one obtains a modified LCA law with a temperature-dependent prefactor:  
\begin{equation}
\label{LCAM}
L(t)= \left(2 a_0 e^{-\beta E_B} t \right)^{1/2} .
\end{equation}

LMV and Paul et al. \cite{ppr04} have also considered systems where $E_B(L)$ has a power-law and logarithmic dependence on $L$, respectively. These usually occur in systems with quenched disorder, where the boundaries of coarsening domains are trapped by disorder sites \cite{hh85,lmv88,sp04}. In the case of power-law barriers, the asymptotic growth law is logarithmic in time \cite{lmv88}. On the other hand, for logarithmic barriers, the growth law is a power-law in time. However, the growth exponent depends on the disorder amplitude and temperature.

\begin{figure}
\centering
\includegraphics[width=0.95\textwidth]{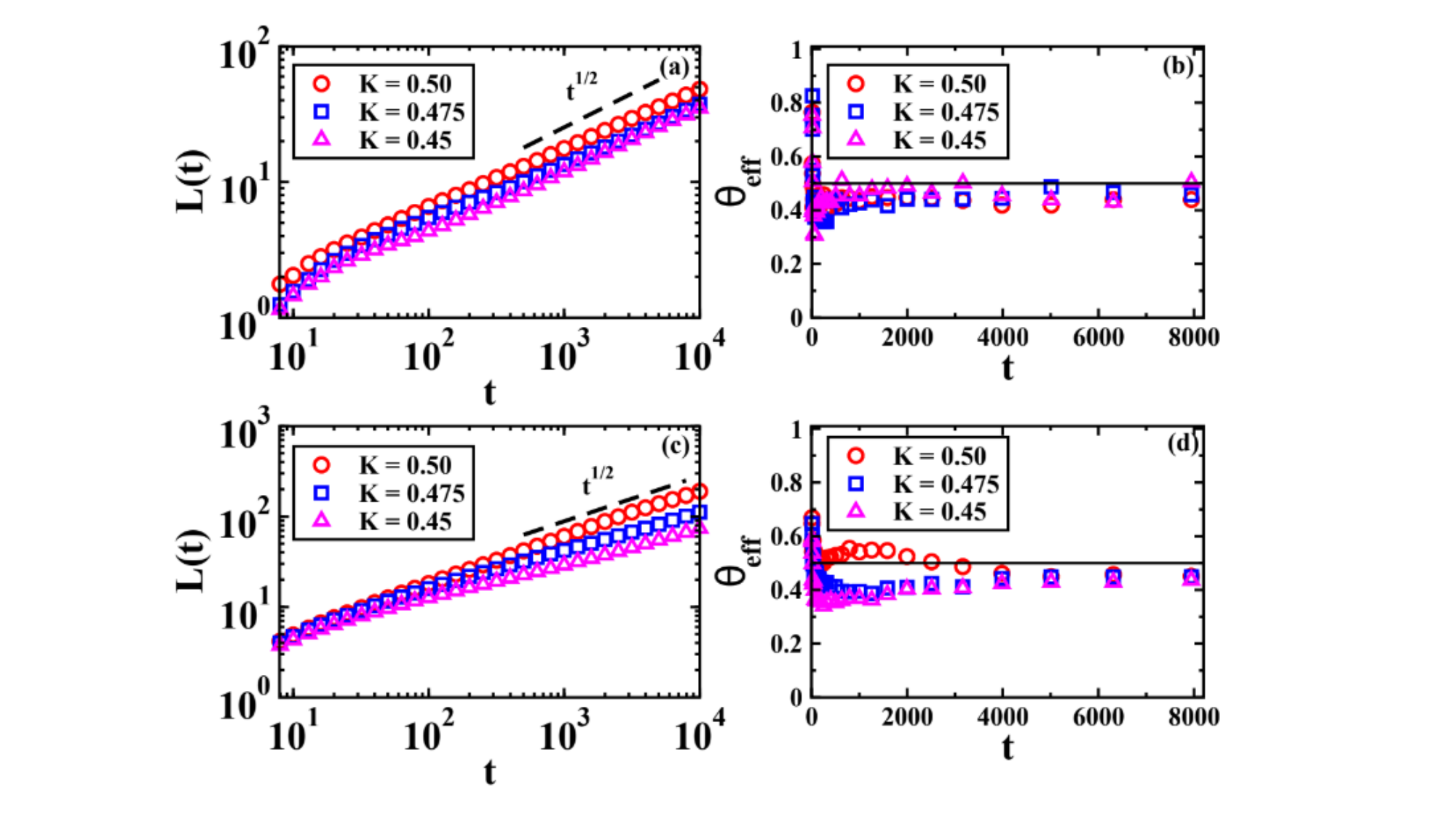}
\caption{(a) Time-dependence of the characteristic length scale, $L(t)$ vs. $t$ on a log-log scale, for the CG. We plot data for different $K$-values. The line labeled $t^{1/2}$ denotes the LCA law. (b) Plot of effective exponent for CG: $\theta_{\rm eff}$ vs. $t$. The horizontal line denotes the LCA exponent, $\theta = 1/2$. (c) Analogous to (a), but for the AF. (d) Analogous to (b), but for the AF.}
\label{f5}
\end{figure}

The plot of $L(t)$ vs. $t$ for the CG is shown in Fig.~\ref{f5}(a). One sees that the growth is slightly slower than the LCA law for all filling factors considered. This is clear from the plot of the effective exponent $\theta_{\rm eff}$ vs. $t$ in Fig.~\ref{f5}(b), where
\beq
\theta_{\rm eff} = \frac{d (\ln L)}{d (\ln t)} .
\eeq
This plot underestimates the LCA exponent ($\theta = 1/2$) for the asymmetric CG. We interpret this as a possible signal of logarithmic growth in the asymptotic regime. Further, the growth is slower for higher asymmetry. We will discuss the nature of the asymptotic growth law at length shortly. The corresponding results for the AF are shown in Figs.~\ref{f5}(c)-(d). These are consistent with the results for the CG. In earlier work, Das et al. \cite{ddp17} have obtained similar results for the AF, via both the kinetic spin model and its phenomenological counterpart (Model C).

\begin{figure}
\centering
\includegraphics[width=0.95\textwidth]{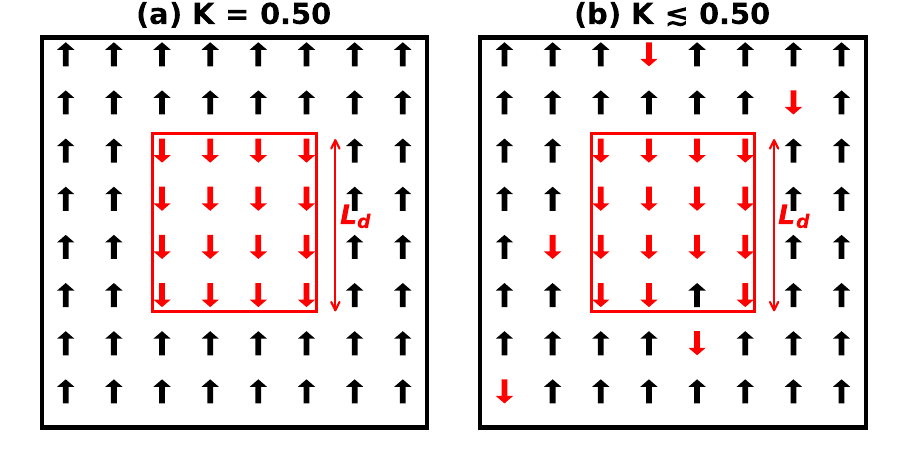}
\caption{Schematic for shrinking of a square droplet (of size $L_d^2$) with $\sigma =-1$ in a background of $\sigma = +1$ in the CG at (a) $K = 0.50$ and (b) $K \lesssim 0.50$. We approximate $\mathcal{H}_{\rm CG}$ by the first two terms in Eq.~(\ref{hsm}). In (a), there is a barrier for flipping the corner spin. The edge subsequently peels without any further barriers. In (b), impurity spins in the background/droplet yield further $L_d$-dependent barriers to the peeling of the edge.}
\label{f6}
\end{figure}

We would like to gain a better understanding of the slowing down of the growth law in the asymmetric case. In this context, following Shore et al. \cite{shs92}, we investigate the time required for shrinking of a droplet on a square lattice. Here, the relevant order parameter is the staggered magnetization $\sigma_i$, so we consider the shrinking of a square droplet ($\sigma_i = -1$) of size $L_d^2$ in a background of $\sigma_i = +1$. This is schematically depicted in Fig.~\ref{f6}(a). If $K \lesssim 0.5$, there will be uniformly scattered impurities (corresponding to excess holes) in the bulk domains -- see Fig.~\ref{f3}. The probability of encountering such an impurity is $p_h = 1-2K$. The corresponding shrinking experiment for $K \lesssim 0.5$ is schematically depicted in Fig.~\ref{f6}(b).

It is convenient to rewrite the first few quadratic terms of $\mathcal{H}_{\rm CG}$ in Eq.~(\ref{Hamiltonian}) in terms of the staggered magnetization:
\beq
\label{hsm}
\mathcal{H}_{\rm CG}\{\sigma_i\} = -J_1 \sum_{nn} \sigma_i \sigma_j +J_2 \sum_{nnn} \sigma_i \sigma_j +J_3 \sum_{nnnn} \sigma_i \sigma_j + \cdots , \quad \sigma_i = \pm 1,
\eeq
where the subscript nnn refers to a sum over next-nearest-neighbor pairs, and so on. In Eq.~(\ref{hsm}),
\beq
J_1 = \frac{e^2}{4\kappa a}, \quad J_2 = \frac{e^2}{4\sqrt{2} \kappa a}, \quad J_3 = \frac{e^2}{8 \kappa a}, \quad \cdots .
\eeq
The argument below is based on the first two terms of the right-hand-side (RHS) of Eq.~(\ref{hsm}) \cite{shs92}, and illustrates the key features which result in formation of energy barriers due to excess holes.

First, consider the case $K=0.5$ in Fig.~\ref{f6}(a). We consider the case where decimation is initiated at the upper left corner. The droplet shrinks by flipping a corner spin with energy barrier $4 J_2$. The subsequent flipping of the edge spins requires zero energy. Thus, the time taken to decimate the square domain is 
\beq
\tau_d = t_0(L_d) \exp \left(\frac{4 J_2}{T} \right) .
\eeq
The decimation of droplets is the elementary process that drives the ordering process. Therefore, we identify $E_B(L)$ in Eq.~(\ref{diff_const}) as $4 J_2$ for the case $K=0.5$. This yields the LCA growth law in Eq.~(\ref{LCAM}).

Next, consider the case $K \lesssim 0.5$, depicted in Fig.~\ref{f6}(b). Again, the decimation is initiated at the upper left corner. The corner flip requires energy $4J_2$ for the case shown in Fig.~\ref{f6}(b). However, in this case, we encounter adjacent impurity spins in the flipping of edge spins. For example, the impurity shown in Fig.~\ref{f6}(b) yields an additional barrier of $4 J_1$. In general, an impurity yields an additional barrier $\alpha_1 J_1 + \alpha_2 J_2$, where the factors $\alpha_1, \alpha_2$ depend on the precise location of the impurity relative to the edge. Therefore, in the process of peeling off the edge, we encounter the overall barrier
\beq
E_B(L_d) \simeq 4 J_2 + \gamma (1-2K) L_d ,
\eeq
where the factor $\gamma$ is an average over individual impurity barriers. Thus, the excess holes give rise to an $L$-dependent barrier in the domain growth process, yielding a {\it Class 3} system in the LMV classification scheme \cite{lmv88}.

This will modify the asymptotic growth law to a logarithmic form. The relevant growth equation is obtained from Eqs.~(\ref{growthEq})-(\ref{diff_const}) as
\beq
\label{grmod}
\frac{dL}{dt} = \frac{a_0}{L} \exp \left[-4\beta J_2 - \beta \gamma (1-2K) L \right] .
\eeq
At early times ($t \ll t_c, L \ll L_c$), we can neglect the second term in the exponent. Then,
\beq
\label{early}
L(t) \simeq \left( 2 a_0 e^{-4 \beta J_2} t \right)^{1/2} , \quad t \ll t_c.
\eeq
At late times ($t \gg t_c, L \gg L_c$), we approximate $L \simeq L_c$ in the prefactor of the RHS of Eq.~(\ref{grmod}). This yields the logarithmic growth law:
\beq
\label{late}
L(t) \simeq \frac{1}{\beta \gamma (1-2K)} \ln \left[ \frac{a_0}{L_c} e^{-4 \beta J_2} \beta \gamma (1-2K) t \right] , \quad t \gg t_c .
\eeq
The crossover time and length scales can be obtained by comparison of Eqs.~(\ref{early}) and (\ref{late}) -- they diverge as $K \rightarrow 1/2^-$.

The corresponding results for AF ordering can be obtained by setting $J_2 = 0$ in Eq.~(\ref{grmod}). A couple of remarks are in order here. Firstly, in the above discussion, we have only considered the first two terms of $\mathcal{H}_{\rm CG}$ in Eq.~(\ref{hsm}). We do not expect the incorporation of higher terms to change the nature of the crossover ($t^{1/2} \rightarrow$ logarithmic), which is already captured in the above discussion. However, higher terms will affect the precise details of the crossover, viz., prefactors of growth laws, etc. Secondly, for the small asymmetries considered in our simulations, we expect the crossover to the logarithmic regime to be substantially delayed. The only signature we see of the asymptotic regime is the under-estimation of the growth exponent from the LCA value, $\theta = 1/2$. Thirdly, it is relevant to point out that there have been several experimental studies \cite{shn88,ruh05} which have reported logarithmic ordering kinetics in asymmetric mixtures. The above arguments provide a theoretical basis for understanding the experimental results.

\section{Summary and Discussion}
\label{s4}

Let us conclude this paper with a summary and discussion of our results. We have studied the domain growth kinetics of a {\it Coulomb glass} (CG) model without any external quenched disorder. The CG is a frustrated system due to the long-range repulsive Coulombic interaction between electrons. The important parameter in our simulations is the filling factor $K$ -- a system with $N$ sites has $K N$ electrons and $(1-K) N$ holes. For $K=0.5$, there are an equal number of holes and electrons in the system. Then, the ground state is a checkerboard where electrons and holes are placed on alternating sites. The CG with $K > 0.5$ is equivalent to the CG with filling fraction $1-K$. Therefore, without any loss of generality, we consider the CG with $K < 0.5$. For $K \lesssim 0.5$, there is a small excess of holes. In this case, the ground state is checkerboard-like with excess holes uniformly distributed. For $K \ll 0.5$, more exotic ground states arise, which we do not study in this paper.

We study the phase ordering kinetics in the CG via a kinetic spin model with long-range antiferromagnetic interactions. For purposes of comparison, we also study domain growth in the nearest-neighbor antiferromagnet (AF). A suitable stochastic kinetics is introduced into the system by connecting it with a heat bath \cite{dp04}. As the number of electrons is conserved, the appropriate microscopic kinetics is {\it Kawasaki spin-exchange kinetics}, where nearest-neighbor spins are stochastically interchanged \cite{pw09}. The relevant order parameter is the {\it staggered magnetization} $\sigma$, which is a nonconserved variable. The kinetics of this variable is coupled to that of the conserved density field, corresponding to {\it Model C} in the Hohenberg-Halperin framework \cite{hh77,ddp17}.

Our simulations of the CG and AF with slight asymmetry yield two important results: \\
(a) For both models, the domain growth morphology is universal for small values of asymmetry. The correlation function of the $\sigma$-field is numerically indistinguishable from that for the symmetric case, $K = 0.5$. Further, the scaling function is in excellent agreement with the Ohta-Jasnow-Kawasaki (OJK) function, which describes nonconserved ordering dynamics in a ferromagnet described by {\it Model A} \cite{hh77}. \\
(b) For both models, the domain growth law is somewhat slower than the {\it Lifshitz-Cahn-Allen} (LCA) law: $L(t) \sim t^{1/2}$. Further, the growth law becomes slower as the asymmetry increases.

To gain a better understanding of the growth law, we study the energy barriers associated with shrinking of a square droplet in a background of the other phase. Our results show that the excess holes (impurity spins in the $\sigma$-domains) present a length-dependent barrier to the peeling process of droplet edges for both the CG and AF. This barrier is proportional to the asymmetry ($1-2K$), and is absent at $K=1/2$. Thus, we expect that both the CG and AF obey the LCA growth law at $K=1/2$. However, for $K \lesssim 1/2$, there is a crossover to a logarithmic law for $t \gg t_c$, where $t_c \rightarrow \infty$ as $(1-2K) \rightarrow 0^+$. The values of $1-2K$ studied here are too small for us to explicitly observe the logarithmic regime. However, we speculate that the under-estimation of the growth exponent from $\theta = 1/2$ in our simulations is a signature of the crossover to the logarithmic regime. Clearly, much longer simulations with larger systems (needed to avoid finite-size effects) are required to clearly resolve this issue. In the context of the long-range interactions considered here, these simulations would be computationally very demanding.

A promising direction for further research is a study of the ordering dynamics when the ground states are more exotic, e.g., striped, BCC, FCC, etc. This is a regime where $K$ is considerably different from 1/2 \cite{rpz13}. Needless to say, these exotic ground states are of great importance for applications in materials science and metallurgy.

\subsubsection*{Acknowledgments}

PB gratefully acknowledges the Kreitman School of Advanced Graduate Studies for financial support. The HPC facility at Ben Gurion University of the Negev is gratefully acknowledged for computational resources. VM acknowledges funding from Science and Engineering Research Board, India under the research grant CRG/2022/004029.

\newpage

\end{document}